\documentclass[preprint, showpacs,preprintnumbers,amsmath,amssymb,nofootinbib]{revtex4}

\usepackage{graphicx}
\usepackage{dcolumn} 
\usepackage{bm}      

\newcommand{\bea}{\begin{eqnarray}}
\newcommand{\eea}{\end{eqnarray}}

\begin{document}
\title{Constraints on a variable  dark energy model with recent observations}
\author{  Puxun Wu\;$^{a,b}$ and  Hongwei Yu\;$^{a,}$\footnote{Corresponding
author}}

\address
{ $^a$Department of Physics and Institute of  Physics,\\ Hunan
Normal University, Changsha, Hunan 410081, China
 \\ $^b$School  of Sciences and Institute of  Physics, Central
South University of Forestry and Technology, Changsha, Hunan 410004,
China}

\begin{abstract}
We place, by the maximum likelihood method,  constraints on a
variable dark energy model with the equation of state
$w=w_0/[1+b\ln (1+z)]^2$ using some recent observational data,
including the new Sne Ia data from the SNLS, the size of baryonic
acoustic oscillation peak from SDSS and the CMB data from WMAP3.
We find that the SNLS data favor models with $w_0$ around  $-1$,
in contrast to the Gold data set which favors a more negative
$w_0$. By combining these three databases, we obtain that
$\Omega_m=0.27_{-0.038}^{+0.036}$, $w_0=-1.11_{-0.30}^{+0.21}$ and
$b=0.31^{+ 0.71}_{-0.31}$ with $\chi^2=110.4$ at the $95\%$
confidence level. Our result suggests that a varying dark energy
model and a crossing of the $w = -1$ line are favored,  and the
present value of the equation of state of dark energy is very
likely less than $-1$.
\end{abstract}


 \maketitle

\section{Introduction}
Since the Hubble diagram of Type Ia Supernovae (Sne Ia)~\cite{Per}
first indicated that the universe is undergoing an accelerating
expansion, many works have been done, trying to explain this
phenomenon. A large number of models for the cosmic acceleration
are proposed by assuming the existence of an energy component with
negative pressure in  the universe, named dark energy, which at
late times dominates the total energy density of the universe and
drives its acceleration of expansion. The simplest candidate of
dark energy is the cosmological constant $\lambda$~\cite{Wein},
with the equation of state $w=p/\rho=-1$. However, the
cosmological constant has to be extremely fine-tuned in order to
induce the observed cosmic expansion,  and this has led many
authors to use scalar fields, such as quintessence~\cite{Peeb},
phantom~\cite{Cald}, quintom~\cite{Quintom} and Chaplygin
gas~\cite{Chaplygin}, as alternative models for dark energy.  On
the other hand there are also some other models where the observed
cosmic acceleration is not driven by dark energy,  such as
modified gravity~\cite{modify}, theories with compactified extra
dimensions~\cite{extra}, DGP model~\cite{GDP} and Cardassian
cosmology~\cite{Cardass} etc.

In order to determine the recent expansion history of our universe,
one can also use a different approach, that is, to assume an
arbitrary parametrization for the equation of state $w(z)$ for dark
energy, where  $z$ is the redshift. The parametrization may not be
motivated by any particular fundamental physical theory and is thus
"model-independent".   It however needs to be designed to give a
good fit to the observational data. The simplest parametrization is
$w = constant$. Some other proposals,  including $w(z) = w_0 + w_1
z$~\cite{line2}, $ w(z) = w_0 + w_1 z/(1 + z)$~\cite{line1} etc, are
also made. Recently Wetterich~\cite{Wette} proposed an interesting
phenomenological parametrization for a variable dark energy, in
which the effective equation of state  is expressed as:
  \bea\label{wz}
 w(z)=\frac{w_0}{[1+b\ln (1+z)]^2}\;,
\eea where  $w_0$ represents the present value of the equation of
state and  $b$ is a positive constant characterizing the change of
$w(z)$ with redshift. Apparently with this $w(z)$ there are three
model parameters ($\Omega_m, w_0, b$) to be determined by the
observations, where $\Omega_m=\rho_m/\rho_c$ represents the
present matter density parameter and $\rho_c=3H^2_0/8\pi G$ is the
present critical density of our universe.  The advantage of this
parametrization  over other parameterizations of a time varying
equation of state, $w(z)$  is that it covers the whole available
redshift range while other parameterizations  proposed before
cover only a restricted range of redshift \cite{line2, line1, wz}.
Later, using Cosmic Microwave Background, Large Scale Structure,
and Sne Ia data, Doran, Karwan and Wetterich~\cite{Doran}
discussed the $w_0$ and the dark energy fraction at very high
redshift $\Omega_d^e$ in this model, and found at the $95\%$
confidence level $w_0<-0.8$ and $\Omega_d^e<0.03$, where
$\Omega_d^e$ and $w_0$ are related to $b$ by $b=-3w_0(\ln
\frac{1-\Omega_d^e}{\Omega_d^e}+\ln\frac{1-\Omega_d^0}{\Omega_d^0})^{-1}$
with $\Omega_d^0$ being the present energy density of dark energy.
Movahed and Rahvar~\cite{Mova} have used the Gold Sna Ia
data~\cite{Gold}, the position of first acoustic peak of the
Cosmic Microwave Background radiation (CMB) and the size of
baryonic acoustic oscillations peak to constrain this model and
obtained at the $2\sigma$ confidence level
$\Omega_m=0.27_{-0.03}^{+0.04}$, $w_0=-1.45_{-2.1}^{+0.65}$ and
$b=1.35^{+ 6.30}_{-1.35}$.

Recently Asitier et al.~\cite{SNLS} released the data of high
redshift supernovae from the Supernova Legacy Survey (SNLS). In
this survey the systematic uncertainties and systematic errors are
reduced. It is worth noting that the SNLS data set is a better
agreement with the WMAP data compared to the Gold Sne Ia
set~\cite{Jassal2006}. Thus, in this paper we will reexamine this
variable dark energy by using the 115 new SNLS Sne Ia data, the
size of baryonic acoustic oscillations peak detected in the
large-scale correlation function of luminous red
galaxies from Sloan Digital Sky Survey (SDSS)~\cite{SDSS} 
and the CMB data obtained from the  three-year WMAP
result~\cite{Wang}. We obtain at a $95\%$ confidence level
$\Omega_m=0.27_{-0.038}^{+0.036}$, $w_0=-1.11_{-0.30}^{+0.21}$ and
$b=0.31^{+ 0.71}_{-0.31}$.

\section{The basic equation}

 By using Eq.~(\ref{wz}) and the equation of energy conservation,
it is easy to obtain the evolution of density of dark
energy~\cite{Wette, Mova}
 \bea\label{rho}
 \rho_d=\rho_{d0}(1+z)^{3(1+\tilde{w}(z))}\;,
\eea
 where $\rho_{d0}$ denotes the present density of
 dark energy and $\tilde{w}(z)=\frac{w_0}{1+b\ln (1+z)}$. Since WMAP observations  strongly indicate that the geometry of our universe is
 spatially flat~\cite{WMAP}, we will ignore the term containing curvature factor in Friedman equation.
  If the radiation components in universe are further
  ignored, we find for the Hubble parameter
  \bea\label{H2}
H^2(z;\Omega_m,w_0)=H_0^2[\Omega_m
(1+z)^3+(1-\Omega_m)(1+z)^{3(1+\tilde{w}(z))}]\;,
 \eea
where  $H_0$ is the present Hubble constant.  Meanwhile one can show
that for a flat universe the Luminosity distance, $d^L$,  can be
expressed as
  \bea\label{dl}
d^L(z,H_0,\Omega_m,w_0)=\frac{c}{H_0}(1+z)\int_0^z\frac{dz'}{E(z',\Omega_m,w_0)}\;.
\eea
 Here $c$ is the velocity  of light  and
 $E(z,\Omega_m,w_0)=H(z;\Omega_m,w_0)/H_0$.

\section{constraints from Sne Ia, SDSS and CMB data }

The the 115 new Sne Ia data includes 44 previously published nearby
Sne Ia and 71 distant Sne Ia released recently by the Supernova
Legacy Survey (SNLS)~\cite{SNLS} which is a planned five year survey
of SNe Ia with $z<1$. Constraints from Sne Ia can be obtained by
fitting the distance modulus $\mu(z)$
\begin{eqnarray}
\mu(z)=5\log_{10}[D^L(z)]+\mathcal{M}\;.
\end{eqnarray}
Here $D^L=H_0d^L$ and $\mathcal{M}=M-5log_{10}(H_0)$, $M$ being the
absolute magnitude of the object.

Recently Eisenstein et al~\cite{SDSS} successfully found the size
of baryonic acoustic oscillation peak using a large spectroscopic
sample of luminous red   galaxy from the SDSS and obtained a
parameter $A$, which is independent of dark energy models and for
a flat universe can be expressed as
\begin{eqnarray}
A=\frac{\sqrt{\Omega_m}}{E(z_1)^{1/3}}\bigg[\frac{1}{z_1}
  \int_0^{z_1}\frac{dz}{E(z)}\bigg]^{2/3}\;,
\end{eqnarray}
 where $z_1 = 0.35$ and $A$ is measured to be $A = 0.469\pm 0.017$.
Using  parameter $A$ we can obtain the constraint on dark energy
models from the SDSS.

For the CMB data, the shift parameter $R$ can be used to constrain
the dark energy  models and it can be expressed as~\cite{shiftR}
 \begin{eqnarray}
 R=\sqrt{\Omega_m}\int_0^{z_r}\frac{dz}{E(z)}\;,
\end{eqnarray}
for a flat universe, where $z_r = 1089$. From the three-year WMAP
result~\cite{WMAP}, the shift parameter is constrained to be  $R =
1.70 \pm 0.03$~\cite{Wang}.

\begin{figure}[htbp]
\includegraphics[width=9cm]{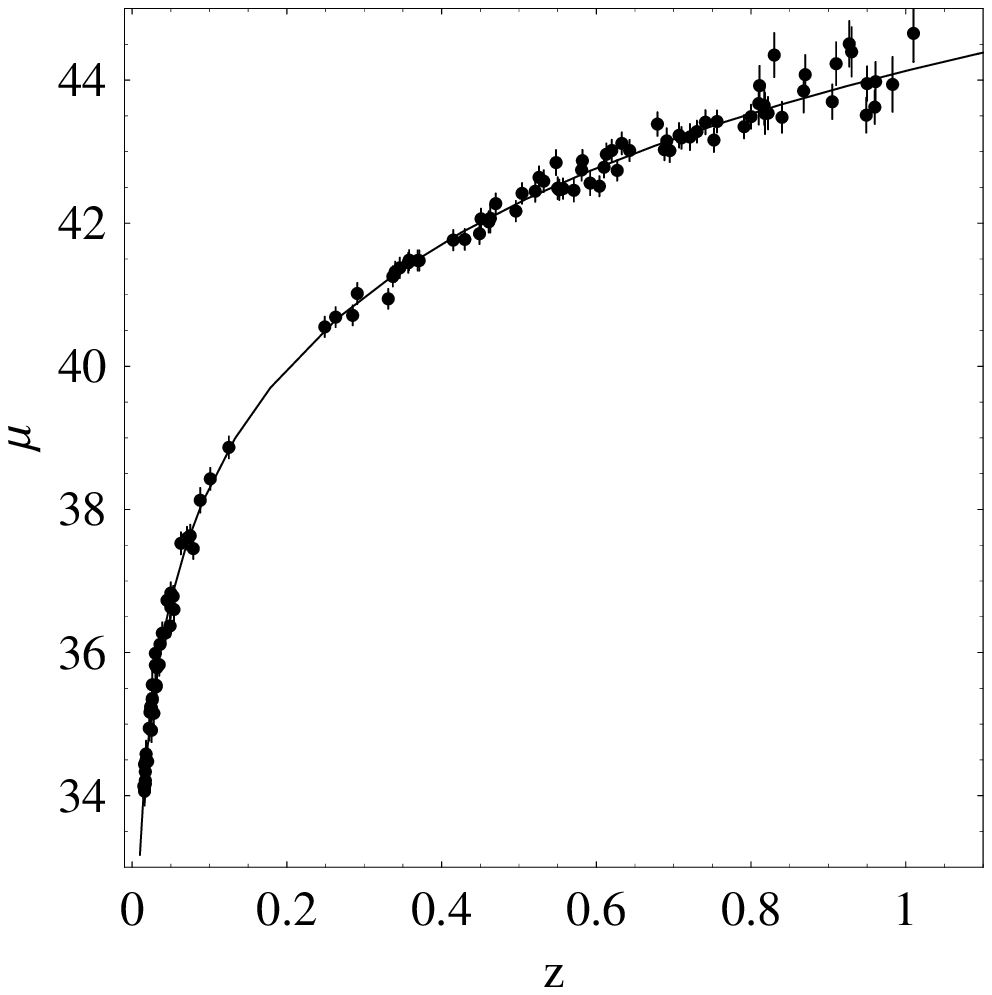}
\caption{\label{Fig1} The  Hubble  diagram for 115 SNLS Sne Ia data
with the best fit parameters $(\Omega_m, w_0, b)=(0.27, -1.11,
0.31)$ obtained from the combination of  SNLS, SDSS and CMB
databases. }
\end{figure}

\begin{figure}[htbp]
\includegraphics[width=10cm]{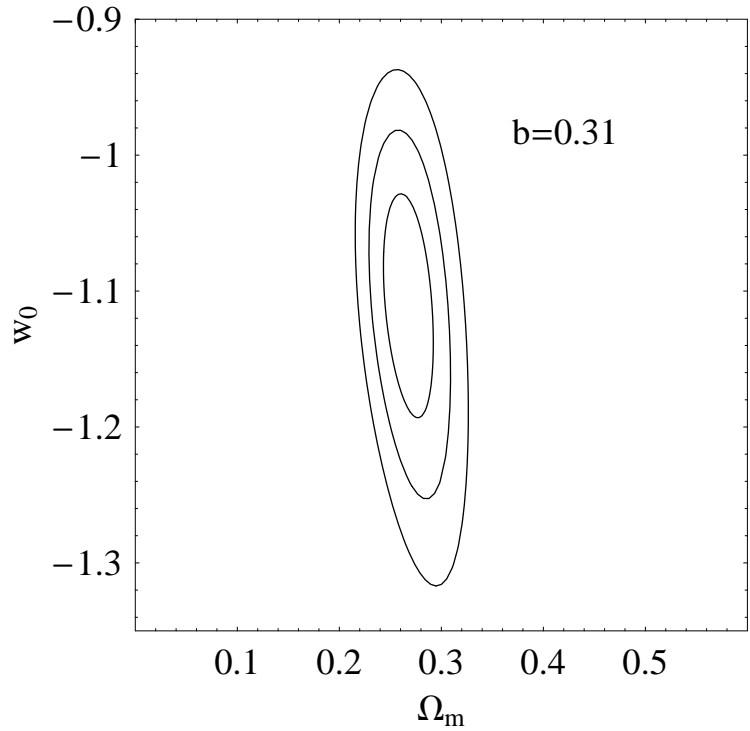}
\caption{\label{Fig2} The $1\sigma$, $2\sigma$ and $3\sigma$
confidence contours for $\Omega_m$ and $w_0$ with $b$ at its best
fit value $0.31$ from the combination of SNLS, SDSS and CMB
databases. }
\end{figure}

\begin{figure}[htbp]
\includegraphics[width=10cm]{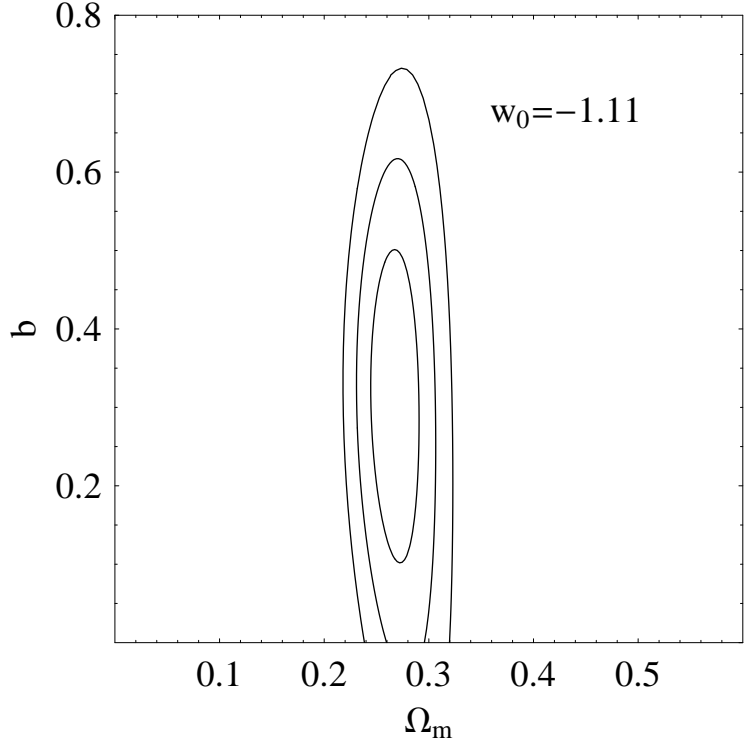}
\caption{\label{Fig3} The $1\sigma$, $2\sigma$ and $3\sigma$
confidence contours for $\Omega_m$ and $b$ with $w_0$ at its best
fit value $-1.11$ from the combination of Gold, SNLS, SDSS and CMB
databases. }
\end{figure}

\begin{figure}[htbp]
\includegraphics[width=10cm]{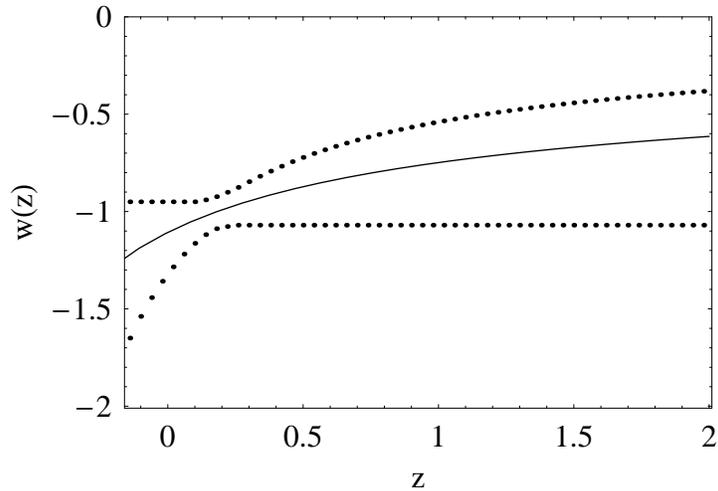}
\caption{\label{Fig4} The behavior of $w(z)$. The solid line plots
$w(z)$ by using the best fit parameters ($w_0=-1.11$,$b=0.31$)
obtained from the SNLS+SDSS+CMB data and the dotted lines are for $1
\sigma$ errors. The solid line shows clearly a varying equation of
state parameter.}
\end{figure}

In order to place limits on model parameters ($\Omega_m, w_0, b$)
with the observation data, we make  use of the maximum likelihood
method, that is, the best fit values for these parameters can be
determined by minimizing
 \begin{eqnarray}
 \chi^2=\Sigma_{i}\frac{[\mu_{obs}(z_i)-\mu(z_i)]^2}{\sigma_i^2}+\frac{(A-0.469)^2}{0.017^2}
 +\frac{(R-1.70)^2}{0.03^2}\;.
 \end{eqnarray}
For the SNLS Sne Ia data set, at a $95.4\%$ confidence level we
obtain $\Omega_m=0.27_{-0.27}^{+0.25}$, $w_0=-1.03_{-1.42}^{+0.46}$
and $b=0.0^{+4.3}$. These are different from the result obtained in
Ref.~\cite{Mova} using  157 Gold Sne Ia data, where $\Omega_m =
0.01^{+0.51}_{-0.01}$, $w_0 =-1.90^{+0.75}_{-3.29}$ and $b =
6.00^{+7.35}_{-6.00}$. Apparently the SNLS data set favors models
with $w_0$ around $-1$ while the Gold set favors a more negative
$w_0$. Meanwhile the best fit value of $b(=0)$ for SNLS data show
that a non-varying dark energy model is favored. However, if
combining the SNLS Sne Ia, SDSS and CMB, we find that at a $95\%$
confidence level
 $\Omega_m=0.27_{-0.038}^{+0.036}$,
$w_0=-1.11_{-0.30}^{+0.21}$ and $b=0.31^{+ 0.71}_{-0.31}$ with
$\chi^2=110.4$. The best fit values show that a variable dark energy
model is favored since $b$ is nonzero and it is very likely  the
present value of the equation of state is less than $-1$. In
Fig.~(\ref{Fig1}) the Hubble Diagram for 115 SNLS Sne Ia data set is
shown with the $(\Omega_m, w_0, b)=(0.27, -1.11, 0.31)$. The contour
plots of $\Omega_{m}$ and $w_0$ by fixing $b$ at its best fit value
$0.31$ are shown in Fig.~(\ref{Fig2}). The contour plots of
$\Omega_{m}$ and $b$ by fixing $w_0$ at its best fit value $-1.11$
are shown in Fig.~(\ref{Fig3}).  In Figs. (\ref{Fig4}) we give the
evolutionary curves $w(z)$ vs $z$ with $1 \sigma$ error bar based on
SNLS Sne Ia+SDSS+CMB data.  This figure shows graphically that SNLS
Sne Ia +SDSS+CMB data favor a varying dark energy model. Comparing
with the results obtained in Ref.~\cite{Mova}, we find that in our
results at the $95\%$ confidence level stronger constraints on $w_0$
and $b$ are obtained. Meanwhile it is easy to see that we also
obtained a stronger constraint on $w_0$ than that obtained in
Ref.~\cite{Doran}.

\section{conclusion}
In this paper we have placed constraints on a parameterized dark
energy model \cite{Wette} using the new SNLS Sne Ia data sets, the
size of the baryonic acoustic oscillation peak from SDSS and the
shift parameter from the CMB observation.  It is found that a
non-varying phantom dark energy model is favored if the SNLS Sne Ia
data set is used in contrast to a varying dark energy when  the Gold
Sne Ia data is utilized ~\cite{Mova}, and different from the case of
the Gold set, the model with $w_0$ around $-1$ is favored by SNLS
data at a $95.4\%$ confidence level. Combing three databases (SNLS
Sne Ia, SDSS and CMB), we obtain a constraint on the model
parameters $(\Omega_m, w_0,b )$ , which suggests that a varying dark
energy model and a crossing of the $w = -1$ line are favored
\cite{cross}, and the present value of the equation of state of dark
energy is very likely less than $-1$.

\begin{acknowledgments}
We would like to thank Z.H. Zhu for his helpful discussions. This
work was supported in part by the National Natural Science
Foundation of China  under Grants No. 10375023 and No. 10575035,
and the Program for NCET under Grant No. 04-0784.
\end{acknowledgments}

\end{document}